\documentclass{ptptex}

\usepackage{graphicx}
\title{%        %You can use \\ for explicit line-break
Tilting Instability in Negative-$\gamma$ Rotating Nuclei%
}

\author{%       %Use \scshape  for the family name
Masayuki \textsc{Matsuzaki}\footnote{E-mail: matsuza@fukuoka-edu.ac.jp}%
}

\inst{%         %Affiliation, neglected when [addenda] or [errata]
Department of Physics, Fukuoka University of Education, \\
Munakata 811-4192, Japan
}
\abst{%         %this abstract is neglected when [addenda] or [errata]
 Based on the cranking model and the random phase approximation, we point 
out that the wobbling excitation on top of the $s$ band in $^{182}$Os is stable 
against angular momentum tilting. 
This is consistent with the general trend that the wobbling excitations in 
$\gamma<0$ rotating nuclei are more stable than those in $\gamma>0$ ones 
found in our previous studies. In higher $N$ isotopes known to be $\gamma$ soft, 
however, a different type of tilting instability is expected. 
Its possible correspondence to the experimental data is also discussed. 
}

\begin{document}

\maketitle

 Symmetry breaking in the nuclear mean field is an analog of second order 
phase transitions 
in infinite systems and is one of the key concepts in the theory of nuclear 
collective motion. According to the general concept of symmetry breaking, 
when one approaches the transition point from the symmetric side, softening 
of collective vibrational mode takes place as a precursor of the phase 
transition. Examples are: 
1) In the spherical to axial shape transition, the $2^+$ quadrupole vibration 
softens, 
2) in the axial to triaxial shape transition, the $\gamma$ vibration softens, 
and 
3) in the normal fluid to superfluid transition, the pair transfer cross section 
increases. 

 Collective rotation of axially symmetric nuclei takes place only about a 
principal axis (usually named the $x$ axis) perpendicular to the symmetry 
axis (the $z$ axis). In triaxially deformed nuclei, however, rotations about 
all three principal axes are possible. Therefore, if triaxiality sets in 
gradually, the angular momentum vector starts to wobble when seen from the 
principal axis frame. Eventually the angular momentum vector tilts permanently 
from the $x$ axis. This regime is called the tilted axis rotation (TAR) in 
contrast to the usual principal axis rotation (PAR). 
Thus, the softening of the wobbling motion is the precursor of symmetry 
breaking from the PAR to the TAR. 
We call this instability of the PAR mean field, caused by the softening 
of the wobbling motion, the tilting instability. 
After this instability, a TAR mean field, in which the signature 
quantum number that is associated with a $\pi$ rotation about the $x$ axis 
is broken, replaces the PAR mean field. 
As shown in Eq.(\ref{dispbm}), the excitation energy of 
the wobbling motion is determined 
by moments of inertia, which are dynamical response of the system to 
rotation from a microscopic viewpoint. Therefore, not only moments of 
inertia depend on $\gamma$ deformation but also $\gamma$ deformation itself 
depends on the rotation frequency 
since rotational alignments of quasiparticles exert shape driving 
effects on the whole system according to their positions in the shell. 

 The small amplitude wobbling motion at high spins was first 
discussed by Bohr and Mottelson~\cite{bm} in terms of a macroscopic 
rotor model with constant moments of inertia. 
Then it was studied microscopically by 
Janssen and Mikhailov~\cite{jm} and Marshalek~\cite{ma} in terms of 
the random phase approximation (RPA) that gives dynamical moments of 
inertia. Since the small amplitude 
wobbling mode has the same quantum number, parity $\pi = +$ 
and signature $\alpha = 1$, as the odd-spin member of the 
$\gamma$ vibrational band, Mikhailov and Janssen~\cite{mj} 
anticipated that it would appear as a high-spin continuation 
of the odd-spin $\gamma$ band. But it has not been clear that 
in which nuclei, at what spins, and with what shapes 
it would appear. Using the RPA, Shimizu and Matsuyanagi~\cite{sm} 
studied Er isotopes with small $|\gamma|$, Matsuzaki~\cite{mm} 
and Shimizu and Matsuzaki~\cite{smm} studied $^{182}$Os with a 
rather large negative $\gamma$ but their correspondence to 
experimental data was not very clear. In 2001, 
{\O}deg{\aa}rd {\it et al.}~\cite{lu1} found an excited triaxial 
superdeformed (TSD) band in $^{163}$Lu and identified it firmly as a 
wobbling band by comparing the observed and theoretical interband 
$E2$ transition rates. These data were investigated in terms of 
a particle-rotor model (PRM) by Hamamoto~\cite{hama} and in terms of 
the RPA by Matsuzaki {\it et al.}~\cite{msmr}. 
In the latter, the calculated dynamical moments of inertia are rotation 
frequency dependent even when the shape of the mean field is fixed. 
This dependence is essential for understanding the observed 
behavior of the excitation energy. In 2002, two-phonon 
wobbling excitations were also observed by Jensen 
{\it et al.}~\cite{2phonon} and their excitation energies show some 
anharmonicity. In Ref.~\citen{msmr}, a numerical example of the softening 
of the wobbling motion in the positive-$\gamma$ nucleus, $^{147}$Gd, was 
presented. Matsuzaki and Ohtsubo~\cite{mo} elucidated that study by examining 
shape change of the potential surface as a function of the tilting angles. 
In that paper it was also discussed that the observed anharmonicity may be a 
signature of the onset of softening. Oi~\cite{oi2} proposed a new model to 
account for this softening. Almehed {\it et al.}~\cite{al} also discussed this. 
Recently Tanabe and Sugawara-Tanabe proposed an approximation method to solve 
the PRM and applied it to the TSD bands~\cite{tan}. Kvasil and Nazmitdinov gave a 
prediction for the wobbling excitations in normal deformed nuclei~\cite{kn} 
by utilizing the sum rule type criterion found in Ref.~\citen{msmf}.

 The excitation energy of the wobbling motion is given, 
as a function of moments of inertia, by~\cite{bm}
\begin{equation}
\hbar\omega_\mathrm{wob}=\hbar\omega_\mathrm{rot}
\sqrt{\frac{\left(\mathcal{J}_x-\mathcal{J}_y\right)
   \left(\mathcal{J}_x-\mathcal{J}_z\right)}
     {\mathcal{J}_y
      \mathcal{J}_z}} \ ,
\label{dispbm}
\end{equation}
where $\omega_\mathrm{rot}$ is the frequency of the main rotation about the 
$x$ axis and $\mathcal{J}$s are moments of inertia about three principal axes. 
This indicates that 
$\mathcal{J}_x>\mathcal{J}_y,\mathcal{J}_z$ or
$\mathcal{J}_x<\mathcal{J}_y,\mathcal{J}_z$ 
must be fulfilled for $\omega_\mathrm{wob}$ to be real. 
The irrotational model moment of inertia is given by 
\begin{equation}
\mathcal{J}_k^\mathrm{irr}\propto\sin^2{(\gamma+\frac{2}{3}\pi k)} ,
\label{irr}
\end{equation}
with $k =$ 1 -- 3 denoting the $x$ -- $z$ principal axes, and its $\gamma$ 
\textit{dependence} is believed to be realistic. When this is taken, 
$-60^\circ<\gamma<0$ for the former or 
$-120^\circ<\gamma<-90^\circ$ or $30^\circ<\gamma<60^\circ$ for the latter 
is required. Since the $\gamma$ deformation of the observed TSD band is 
$\gamma\sim+20^\circ$, another mechanism is necessary for the wobbling 
excitation to exist. It was  found in Ref.~\citen{msmr} and elucidated in 
Ref.~\citen{msmf} that the alignment of the last odd quasiproton brings an 
additional contribution to $\mathcal{J}_x$ and consequently makes 
$\mathcal{J}_x>\mathcal{J}_y$ in place of 
$\mathcal{J}_x<\mathcal{J}_y$ in the irrotational-like behavior. 
But the smallness of $\mathcal{J}_x-\mathcal{J}_y$ implies fragility of the 
excitation. 

 The negative-$\gamma$ collective rotation, $-60^\circ<\gamma<0$, is 
expected to occur prevailingly. However, it looks difficult to excite a 
wobbling mode on the ground band of even-even nuclei because of 
$\mathcal{J}_x\sim\mathcal{J}_y$ in those cases (see Ref.~\citen{mm}). 
Therefore, the following three conditions are desirable for the wobbling 
excitation to exist: 
1) $-60^\circ<\gamma<0$, 
2) $|\gamma|$ is not small, 
and 
3) existence of aligned quasiparticle(s) that makes $\mathcal{J}_x$ larger. 
From these conditions, we chose the $s$ band of $^{182}$Os 
as a representative in Refs.~\citen{mm,smm}. 
We concluded that a wobbling excitation exists on top of the $s$ band of 
$^{182}$Os. Recently, Hashimoto and Horibata~\cite{hh} presented the opposite 
conclusion.
Here we briefly comment on their work before proceeding to the main discussion of 
this paper. They recently reported a renewed three-dimensional cranking 
calculation for $^{182}$Os paying attention to the stability of the $s$ band 
based on their previous calculation~\cite{hori}. They concluded that the 
wobbling excitation on the $s$ band does not exist; this contradicts our 
previous calculation~\cite{mm,smm}. A close looking into their works leads 
one to find that the character of the $s$ band is different between theirs and 
ours. Although not stated in Ref.~\citen{hh}, it was reported in 
Ref.~\citen{hori} that their $s$ band consists of two aligned 
\textit{quasiprotons}. Their low-$\Omega$ $h_{9/2}$ character would lead to a 
positive-$\gamma$ shape. Note that their convention for the sign of $\gamma$ 
is opposite to the Lund convention adopted here. As stated above, 
wobbling excitations in positive-$\gamma$ nuclei are fragile. Although our 
calculation adopted fixed mean field parameters, we conformed to the experimental 
information that suggests the $s$ band consists of two aligned $i_{13/2}$ 
\textit{quasineutrons}~\cite{os182}. Since the Fermi surface is located at a 
high position in the $i_{13/2}$ shell, the alignment leads to a negative-$\gamma$ 
shape. As discussed above, the wobbling excitations on negative-$\gamma$ 
quasiparticle aligned configurations are rather stable. This is the reason why 
the conclusions of Hashimoto and Horibata and ours are different. The 
collective excitation on the $g$ band is expected or exists in both calculations, 
but in our calculation it is $\gamma$ vibration-like rather than 
wobbling-like (see Ref.~\citen{mj}). 

 Now we proceed to present a numerical example of another type of tilting 
instability in negative-$\gamma$ rotating nuclei, 
different from that in positive-$\gamma$ cases discussed in our previous 
works~\cite{msmr,mo}, although wobbling excitations 
are stable in many negative-$\gamma$ cases when it exists. 
The meaning of ``different" is elucidated later. First, we review 
our model briefly.  We begin with a one-body Hamiltonian in the rotating frame, 
\begin{gather}
h'=h-\hbar\omega_\mathrm{rot}J_x , \\
h=h_\mathrm{Nil}-\Delta_\tau (P_\tau^\dagger+P_\tau)
                   -\lambda_\tau N_\tau , \label{hsp} \\
h_\mathrm{Nil}=\frac{\mathbf{p}^2}{2M}
                +\frac{1}{2}M(\omega_x^2 x^2 + \omega_y^2 y^2 + \omega_z^2 z^2)
                +v_{ls} \mathbf{l\cdot s} 
                +v_{ll} (\mathbf{l}^2 - \langle\mathbf{l}^2\rangle_{N_\mathrm{osc}}) .
                \label{hnil}
\end{gather}
In Eq.(\ref{hsp}), $\tau = 1$ and 2 stand for neutron and proton, respectively, 
and chemical potentials $\lambda_\tau$ are determined so as to give correct average 
particle numbers $\langle N_\tau \rangle$. 
The oscillator frequencies in Eq.(\ref{hnil}) 
are related to the quadrupole deformation parameters $\epsilon_2$ and $\gamma$ 
in the usual way. 
They are treated as parameters as well as pairing gaps $\Delta_\tau$. 
The orbital angular momentum $\mathbf{l}$ in Eq.(\ref{hnil}) is defined in the 
singly stretched coordinates $x_k' = \sqrt{\frac{\omega_k}{\omega_0}}x_k$ 
and the corresponding momenta, with $k =$ 1 -- 3 denoting $x$ -- $z$. 
Since $h'$ conserves parity $\pi$ and signature $\alpha$, nuclear states can be 
labeled by them.  We perform the RPA to the residual pairing plus 
doubly stretched quadrupole-quadrupole ($Q'' \cdot Q''$) interaction between 
quasiparticles. Since we are interested in the wobbling motion that has a 
definite signature quantum number, $\alpha = 1$, 
only two components out of five of the $Q'' \cdot Q''$ interaction are relevant. 
They are given by 
\begin{equation}
H_\mathrm{int}^{(-)}=-\frac{1}{2}\sum_{K=1,2} \kappa_K^{(-)} Q_K''^{(-)\dagger} Q_K''^{(-)} ,
\end{equation}
where the doubly stretched quadrupole operators are defined by 
\begin{equation}
Q_K''=Q_K(x_k\rightarrow x_k'' = \frac{\omega_k}{\omega_0}x_k) ,
\end{equation}
and those with good signature are 
\begin{equation}
Q_K^{(\pm)}=\frac{1}{\sqrt{2(1+\delta_{K0})}}\left(Q_K \pm Q_{-K}\right) .
\end{equation}
The residual pairing interaction does not contribute because $P_\tau$ is an 
operator with $\alpha = 0$. 
The equation of motion, 
\begin{equation}
\left[h'+H_\mathrm{int}^{(-)},X_n^\dagger\right]_\mathrm{RPA}
=\hbar\omega_n X_n^\dagger ,
\end{equation}
for the eigenmode 
\begin{equation}
X_n^\dagger=\sum_{\mu<\nu}^{(\alpha=\pm 1/2)}
\Big(\psi_n(\mu\nu)a_\mu^\dagger a_\nu^\dagger 
+\varphi_n(\mu\nu)a_\nu a_\mu\Big)
\end{equation}
leads to a pair of coupled equations for the transition amplitudes 
\begin{equation}
T_{K,n}=\left\langle\left[Q_K^{(-)},X_n^\dagger\right]\right\rangle .
\end{equation}
Then, by assuming $\gamma \neq 0$, this can be cast~\cite{ma} into the form 
\begin{gather}
(\omega_n^2-\omega_\mathrm{rot}^2)\left[\omega_n^2-\omega_\mathrm{rot}^2
\frac{\left(\mathcal{J}_x-\mathcal{J}_y^\mathrm{(eff)}(\omega_n)\right)
   \left(\mathcal{J}_x-\mathcal{J}_z^\mathrm{(eff)}(\omega_n)\right)}
{\mathcal{J}_y^\mathrm{(eff)}(\omega_n)\mathcal{J}_z^\mathrm{(eff)}(\omega_n)}
                                  \right]=0 .
\label{disp2}
\end{gather}
This expression proves that the spurious mode ($\omega_n=\omega_\mathrm{rot}$; 
not a real intrinsic excitation but a rotation as a whole) given by the first 
factor and all normal modes given by the second are decoupled from each other. 
Here $\mathcal{J}_x = \langle J_x \rangle/\omega_\mathrm{rot}$ as usual and the 
detailed expressions of $\mathcal{J}_{y,z}^\mathrm{(eff)}(\omega_n)$ are given in 
Refs.~\citen{ma,mm,smm}. Among normal modes, one obtains 
\begin{equation}
\omega_\mathrm{wob}=\omega_\mathrm{rot}
\sqrt{\frac{\left(\mathcal{J}_x-\mathcal{J}_y^\mathrm{(eff)}(\omega_\mathrm{wob})\right)
   \left(\mathcal{J}_x-\mathcal{J}_z^\mathrm{(eff)}(\omega_\mathrm{wob})\right)}
     {\mathcal{J}_y^\mathrm{(eff)}(\omega_\mathrm{wob})
      \mathcal{J}_z^\mathrm{(eff)}(\omega_\mathrm{wob})}} ,
\label{disp}
\end{equation}
by putting $\omega_n=\omega_\mathrm{wob}$. 
Note that this gives a real excitation only when the argument of the square root 
is positive and it is non-trivial whether a collective solution appears or not. 
Evidently this coincides with the form (\ref{dispbm}) derived by Bohr and Mottelson 
in a rotor model~\cite{bm} and known in classical mechanics~\cite{landau}. 
Further, this makes it possible to describe the mechanism of the tilting instability 
in terms of the dynamical moments of inertia. The wobbling angles 
that measure the amplitude of the vibrational motion of the angular momentum vector 
around the $x$ axis are defined by 
\begin{gather}
\theta_\mathrm{wob}=\tan^{-1}
{\frac{\sqrt{\vert J_y^\mathrm{(PA)}(\omega_\mathrm{wob})\vert^2
            +\vert J_z^\mathrm{(PA)}(\omega_\mathrm{wob})\vert^2}}
      {\langle J_x^\mathrm{(PA)}\rangle}} , \\
\varphi_\mathrm{wob}=\tan^{-1}
\Bigg|\frac{J_z^\mathrm{(PA)}(\omega_\mathrm{wob})}{J_y^\mathrm{(PA)}(\omega_\mathrm{wob})}\Bigg| ,
\end{gather}
with (PA) denoting the principal axis frame. 
The PA components of the angular momentum vector are defined by 
\begin{gather}
\langle J_x^{\rm (PA)}\rangle=\langle J_x\rangle, \\
iJ_y^{\rm (PA)}=iJ_y-\frac{\langle J_x\rangle}{2\langle Q_2^{(+)}\rangle}Q_2^{(-)}, \\
J_z^{\rm (PA)}=J_z-
\frac{\langle J_x\rangle}{\sqrt{3}\langle Q_0^{(+)}\rangle-\langle Q_2^{(+)}\rangle}Q_1^{(-)}, 
\end{gather}
in terms of the RPA matrix elements of their uniformly rotating frame components 
usually calculated in the cranking model~\cite{ma,mm,smm}, 
because the PA frame is determined by diagonalizing the quadrupole tensor 
$Q_K^\mathrm{(PA)}$~\cite{ma,ko}. 

\begin{figure}[htbp]
\begin{center}
 \includegraphics[width=8cm]{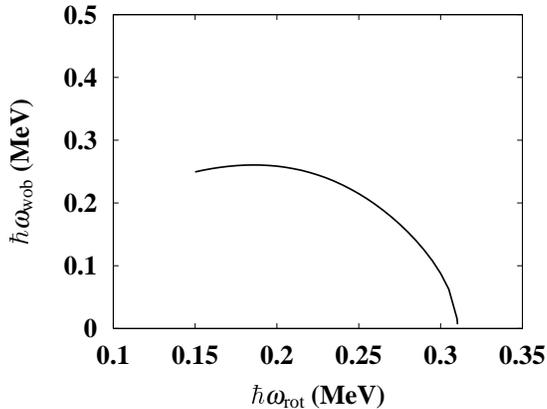} 
\end{center}
 \caption{Rotation frequency dependence of the wobbling excitation energy 
on the $s$ band of $^{186}$Os. \label{fig1}}
\end{figure}

\begin{figure}[htbp]
\begin{center}
 \includegraphics[width=8cm]{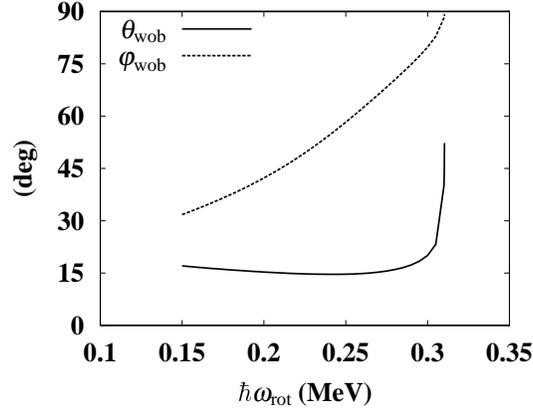} 
\end{center}
 \caption{Rotation frequency dependence of the wobbling angles. \label{fig2}}
\end{figure}

\begin{figure}[htbp]
\begin{center}
 \includegraphics[width=8cm]{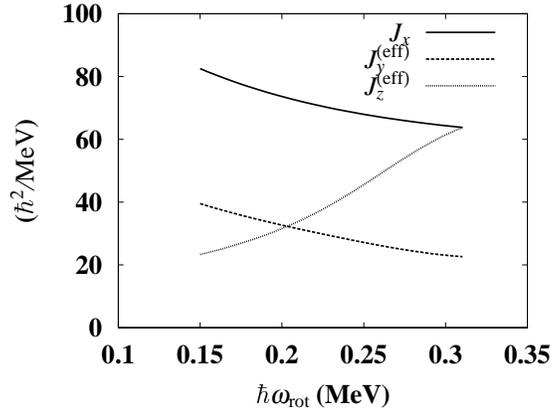} 
\end{center}
 \caption{Rotation frequency dependence of the moments of inertia. \label{fig3}}
\end{figure}

 We choose $^{186}$Os, bearing possible correspondence to the experimental data 
in mind. 
The $s$ band consists of $(\nu i_{13/2})^2$. 
In this calculation, we concentrate on the direct rotational effect by ignoring 
the effect of the possible rotational shape change. 
The adopted mean field parameters are 
$\epsilon_2 = 0.205$, $\gamma= -32^\circ$, and $\Delta_n = \Delta_p = 0.4$ MeV. 
Calculations are performed in the model space of five major shells; 
$N_\mathrm{osc} =$ 3 -- 7 for neutrons and 2 -- 6 for protons. 
The strengths of the $\mathbf{l\cdot s}$ and $\mathbf{l}^2$ potentials are 
taken from Ref.~\citen{br}. Figure~\ref{fig1} reports the excitation energy 
$\hbar\omega_\mathrm{wob}$ in the rotating frame. Decrease of this quantity 
signals the instability of the principal axis rotating $s$ band that 
supports the small amplitude wobbling excitation. Figure~\ref{fig2} graphs 
the wobbling angles $\theta_\mathrm{wob}$ and $\varphi_\mathrm{wob}$. 
While the angular momentum vector wobbles around the $x$ axis with 
$\theta_\mathrm{wob}\simeq15^\circ$ up to just below the instability point, 
$\varphi_\mathrm{wob}$ increases gradually. 
This means that the $z$ component increases gradually. Eventually at the 
instability point the angles look to reach $\theta_\mathrm{wob}>45^\circ$ 
and $\varphi_\mathrm{wob}=90^\circ$, that is, the angular momentum vector 
tilts to the $x-z$ plane. 
Although the present calculation can not go beyond the instability point, 
a numerical example of the correspondence between the instability of the 
PAR and the TAR that follows it was presented in Ref.~\citen{mo}. 
More direct information about the shape that the system would favor 
can be obtained from the moments of 
inertia shown in Fig.~\ref{fig3}. This figure shows that 
$\mathcal{J}_x=\mathcal{J}_z$ is realized at the instability point; this is 
a \textit{different} type of tilting instability from that observed in $\gamma>0$ 
nuclei that is caused by $\mathcal{J}_x=\mathcal{J}_y$. 
Here we elucidate the meaning of ``\textit{different} type". 
The instability brought about by $\mathcal{J}_x=\mathcal{J}_y$ 
discussed in Refs.~\citen{msmr,mo} and that by $\mathcal{J}_x=\mathcal{J}_z$ 
discussed here are similar in the sense that the energy costs 
of rotations about two different axes coincide. 
But here we base our discussion on the physical picture that $\gamma>0$ and 
$\gamma<0$ are different rotation schemes and, according to the reason discussed 
above, at $\gamma>0$, $\omega_\mathrm{wob}$ can not be real without aligned 
quasiparticle that makes $\mathcal{J}_x$ larger, in contrast, at $\gamma<0$, 
$\omega_\mathrm{wob}$ can be real without it. 
Note that nothing peculiar happens at $\mathcal{J}_y=\mathcal{J}_z$ because 
the instability is given by zeros of Eq.~(\ref{disp}). 
Although selfconsistent shape change is beyond the scope of the present simple-minded 
calculation, $\mathcal{J}_x=\mathcal{J}_z$ may indicate that either a TAR 
($\mathcal{J}_y\neq0$) or another PAR, that is, an oblate collective rotation 
($\mathcal{J}_y=0$ for the irrotational rotor), would be favored. 
A possibility of oblate collective rotation was first discussed by Hilton and 
Mang~\cite{hil} for $^{180}$Hf, and very recently by Walker and Xu~\cite{wx} 
and Sun et al.~\cite{sun} for $^{190}$W. 
In the present case, $\mathcal{J}_y$ is decreasing but not 0. Therefore, it is 
natural to regard the rotation scheme just after the instability as a TAR. 

Although the quantitative criterion for the occurrence of the instability is beyond 
the scope of the present calculation, we confirmed that the instability 
occurs at lower rotation frequency for smaller $\epsilon_2$ or larger $N$. 
These results point to consistency with the 
$N$ dependence of the $\gamma$ softness in this mass region seen in the 
quadrupole deformation~\cite{cl}, the excitation energy of the $\gamma$ 
vibration~\cite{bm}, and the high-$K$ isomerism~\cite{whe}. 

 Finally we mention possible correspondence to the observed data. 
In Ref.~\citen{os186}, Balabanski \textit{et al}. reported an anomalous termination 
of the yrast band of $^{186}$Os at 18$^+$. 
According to their calculation, the $(\nu i_{13/2})^2$ alignment drives the shape 
to $\gamma\simeq-30^\circ$ before this termination. Actually the mean field 
parameters of the present calculation were chosen conforming to this. 
As for the termination itself, 
they discussed using a total Routhian surface calculation that it is related to 
a further shape change in the $\gamma$ direction. 
Later Wheldon \textit{et al}.~\cite{whe} discussed that it does not terminate. 
Aside from the different conclusions about the fate of the higher spin states, 
the yrast band changes its character at 14$^+$ in both studies. 
Wheldon \textit{et al}.~\cite{whe} concluded that this is caused by the crossing 
with the $10^+$ band that is tilted (``$t$ band"). 
Since the main component of the high spin part of 
the ground state (gs) band is thought to be a PAR triaxial $(\nu i_{13/2})^2$ $s$ 
band~\cite{wy}, the observed crossing is ascribed to the instability of the PAR 
mean field qualitatively. On the other hand, since the $14^+$ and the $12^+$ 
members of the gs band correspond to $\hbar\omega_\mathrm{rot}=$ 0.389 MeV 
and 0.356 MeV, respectively, the observed crossing takes place between them. 
Therefore, quantitative correspondence with the present calculation 
in which it takes place at around $\hbar\omega_\mathrm{rot}=$ 0.310 MeV 
is insufficient. 

 To summarize, in this paper, first we have pointed out that the wobbling 
excitations on $\gamma<0$ quasiparticle aligned bands are expected to be 
more stable than those on $\gamma>0$ ones as found in our previous studies. 
In relation to this, we have clarified the reason for the 
different conclusions about the existence of the wobbling excitation on top of 
the $s$ band of $^{182}$Os between the recent work of Hashimoto and 
Horibata~\cite{hh} and ours. 
Second, we have discussed, in spite of this, that the wobbling excitation in 
$\gamma<0$ nuclei can become unstable by presenting a numerical example, although the 
quantitative criterion for the occurrence of this type of tilting instability 
is deferred to more elaborate calculations. 
Possible correspondence of this example to the experimental data is also discussed.

%%%%%%%%%%%%%%%%%%%%%%%%%%%%%%%%%%%%%%%%%%%%%%%%%%%%%%%%%%%%%
% Some macros are available for the bibliography:
%  o for general use
%    \JL : general journals                 \andvol : Vol (Year) Page
%  o for individual journal 
%    \AJ   : Astrophys. J.           \NC         : Nuovo Cim.
%    \ANN  : Ann. of Phys.           \NPA, \NPB  : Nucl. Phys. [A,B]
%    \CMP  : Commun. Math. Phys.     \PLA, \PLB  : Phys. Lett. [A,B]
%    \IJMP : Int. J. Mod. Phys.      \PRA - \PRE : Phys. Rev. [A-E]     
%    \JHEP : J. High Energy Phys.    \PRL        : Phys. Rev. Lett.
%    \JMP  : J. Math. Phys.          \PRP        : Phys. Rep.
%    \JP   : J. of Phys.             \PTP        : Prog. Theor. Phys.     
%    \JPSJ : J. Phys. Soc. Jpn.      \PTPS       : Prog. Theor. Phys. Suppl.
% Usage:
%  \PRD{45,1990,345}          ==> Phys.~Rev.\ \textbf{D45} (1990), 345
%  \JL{Nature,418,2002,123}   ==> Nature \textbf{418} (2002), 123
%  \andvol{B123,1995,1020}    ==> \textbf{B123} (1995), 1020
%%%%%%%%%%%%%%%%%%%%%%%%%%%%%%%%%%%%%%%%%%%%%%%%%%%%%%%%%%%%%

\end{document}